\documentclass[sigconf]{acmart}

\usepackage{pgfplots}
\usepackage{pgfplotstable}
\pgfplotsset{compat=newest}
\usepackage{graphicx}
\usepackage{cleveref}

\usepackage{amssymb} 
\usepackage{makecell}
\usepackage{multirow}
\usepackage{url}
\AtBeginDocument{%
  }

\copyrightyear{2025}
\acmYear{2025}
\setcopyright{acmlicensed}
\acmConference[CIKM '25]{Proceedings of the 34th
ACM International Conference on Information and Knowledge
Management}{November 10--14, 2025}{Seoul, Republic of Korea}
\acmBooktitle{Proceedings of the 34th ACM International Conference on
Information and Knowledge Management (CIKM '25), November 10--14, 2025,
Seoul, Republic of Korea}
\acmDOI{10.1145/3746252.3761636}
\acmISBN{979-8-4007-2040-6/2025/11}

\begin{document}

\title{CSMD: Curated Multimodal Dataset for Chinese Stock Analysis}


\author{Yu Liu}
\orcid{0009-0004-8537-2256}
\authornote{Yu Liu and Zhuoying Li contributed equally to this work.}
\affiliation{
  \institution{East China Normal University}
  \city{Shanghai}
  \country{China}
}
\email{yuliu@stu.ecnu.edu.cn}

\author{Zhuoying Li}
\authornotemark[1] 
\orcid{0009-0006-9970-6792}
\affiliation{
  \institution{East China Normal University}
 \city{Shanghai}
 \country{China}
}
\email{zhuoyingli@stu.ecnu.edu.cn}

\author{Ruifeng Yang}
\orcid{0009-0009-4878-1727}
\affiliation{%
  \institution{East China Normal University}
  \city{Shanghai}
  \country{China}
}
\email{rfyang@stu.ecnu.edu.cn}

\author{Fengran Mo}
\orcid{0000-0002-0838-6994}
\affiliation{%
  \institution{University of Montreal}
  \city{Montreal}
  \country{Canada}
}
\email{fengran.mo@umontreal.ca}

\author{Cen Chen}
\authornote{Corresponding Author.}
\orcid{0000-0003-0325-1705}
\affiliation{%
  \institution{East China Normal University}
  \city{Shanghai}
  \country{China}
}
\email{cenchen@dase.ecnu.edu.cn}


\begin{abstract}
The stock market is a complex and dynamic system, where it is non-trivial for researchers and practitioners to uncover underlying patterns and forecast stock movements.
The existing studies for stock market analysis rely on leveraging various types of information to extract useful factors, which are highly conditional on the quality of the data used.
However, the currently available resources are mainly based on the U.S. stock market in English, which is inapplicable to adapt to other countries.
To address these issues, we propose \textbf{CSMD}, a multimodal dataset curated specifically for analyzing the Chinese stock market with meticulous processing for validated quality.
In addition, we develop a lightweight and user-friendly framework \textbf{LightQuant} for researchers and practitioners with expertise in financial domains. Experimental results on top of our datasets and framework with various backbone models demonstrate their effectiveness compared with using existing datasets.
The datasets and code are publicly available at the link: \url{https://github.com/ECNU-CILAB/LightQuant}.
\end{abstract}

\ccsdesc[500]{Information systems~Information integration}

\keywords{Multimodal datasets, Chinese datasets, Stock movement prediction}


\maketitle

\section{Introduction}
The stock market is a complex and dynamic system shaped by a wide range of economic, financial, and psychological factors. While researchers aim to uncover underlying patterns to better understand market behavior, practitioners focus on forecasting stock movements to guide investment decisions \cite{adam2016stock}. Traditional stock market analysis relies mainly on two approaches: fundamental analysis and technical analysis \cite{antoniou1997technical}. Fundamental analysis focuses on macroeconomic and financial data, which often suffers from time lags and depends on subjective human interpretations. Conversely, technical analysis leverages price data, which is heavily impacted by the noisy information and short-term fluctuations from the market, thereby making it difficult to achieve satisfactory accuracy for future stock trend prediction. 
As a result, it is intuitive to incorporate both price information and financial textual data as multi-modality information (e.g., including stock prices and related news texts) to facilitate the modeling for predicting stock trends \cite{nassirtoussi2015text, Nassirtoussi2015AMD, tang2022survey}.

To explore cross-modal relationships between prices and text, prior studies have proposed models such as StockNet \cite{xu2018stock}, HAN \cite{hu2018listening}, and CMIN \cite{luo2023causality}, which leverage complementary information across modalities. However, applying multimodal approaches to stock prediction remains challenging. First, most publicly available datasets focus on the U.S. market and English-language sources, limiting their applicability to markets like China, where financial news and social signals are in Chinese and relevant data are scarce~\cite{huang2024survey}.
Second, financial texts are often polluted by noise, irrelevant content, and low-quality information, which complicates data preprocessing and raises concerns about the quality of the processed data \cite{mao2023discovering}. Furthermore, algorithm researchers spend a significant amount of time on tedious tasks such as dataset selection, data processing, factor selection, baseline model construction, and backtesting program development, which reduces the time allocated for strategy development. Moreover, existing open-source backtesting frameworks are functionally complex and have a high learning curve, making them intimidating for fresh developers.

To address these challenges, we propose \textbf{CSMD}, a multimodal dataset curated specifically for analyzing the Chinese stock market, 
which successfully fills the gap in the current open source datasets that lack the latest time-aligned price and news text data. 
Our CSMD dataset contains the latest price data of China's major stock index components, including two specifications, SSE 50 Index\footnote{\url{https://en.wikipedia.org/wiki/SSE_50_Index}} 
and CSI 300 Index\footnote{\url{https://en.wikipedia.org/wiki/CSI_300_Index}}. 
Targeting multimodal information, we collect financial news aligned with price data, denoised and enhanced via LLMs, to support market analysis from multiple perspectives. Furthermore, we develop \textbf{LightQuant}, a lightweight and user-friendly simulation framework that integrates data processing and backtesting, enabling researchers to conduct strategy evaluation with simple function calls while ensuring comprehensive functionality.
The contributions of this paper are summarized as follows:
\begin{itemize}
    \item We introduce \textbf{CSMD}, a novel multimodal dataset for the Chinese stock market that encompasses stock prices and financial news texts, which is publicly available for research. 
    \item We utilize large language models guided by sequential knowledge prompts to extract news-based factors, which enhance textual relevance, human interpretability, and the explanatory power regarding stock movements.
    \item We develop a lightweight and user-friendly simulated trading backtesting framework, \textbf{LightQuant}, on which extensive experiments demonstrate the effectiveness of CSMD for analyzing the Chinese stock market.
\end{itemize}

\section{Related Work}
Predicting stock price movements is a fundamental challenge in the finance community \cite{de2018advances}.
From early statistical analysis to the development of machine learning and deep learning methods, stock data has always been at the core of analysis.
Previous works mainly focused on using single price data \cite{tang2022survey}, and researchers selected influential indices in different markets as entry points, such as the S\&P 500 and NASDAQ 100 in the US stock market, and the CSI 300 and CSI 100 in the Chinese stock market.

Recently, the development of large language models (LLMs) and multimodal techniques enables the developed models with a powerful capacity to understand the information in different modalities \cite{naveed2023comprehensive, wang2025finsage}.
As a result, existing studies have curated various multimodal datasets, including ACL18 \cite{xu2018stock}, CIKM18 \cite{wu2018hybrid}, BIGDATA22 \cite{soun2022accurate}, and CMIN-US \cite{luo2023causality}, primarily targeting the U.S. stock market, to facilitate multimodal modeling techniques. 
To the best of our knowledge, only one multimodal Chinese stock market dataset, CMIN-CN \cite{luo2023causality}, is publicly available.
The scarcity of data limited the model development for Chinese stock market prediction.
To this end, we construct a high-quality multimodal dataset tailored specifically for the Chinese stock market. 

\section{Dataset Construction}
In this section, we introduce the construction process of our multimodal dataset, CSMD, for Chinese stock analysis, which involves financial news collection, factor extraction enhancement, and data quality validation. In addition to textual data, we collect corresponding price information from Baostock\footnote{\url{www.baostock.com}}.

\subsection{Financial News Data Collection}



All data are collected from \textit{Securities Times}\footnote{http://www.stcn.com/}, a leading 
financial media with extensive coverage and significant influence in China's financial sector. Its authoritative content and broad reach ensure the credibility and reliability of the collected news. By sourcing directly from this reputable outlet, we guarantee the high quality and authenticity of the textual data, forming a solid foundation for subsequent analysis. 
Note that, following licensing and regional policy considerations, our data collection exclusively includes publicly accessible content that requires no subscription or paid access. This approach ensures compliance with legal and ethical standards, effectively mitigating potential conflicts of interest and respecting intellectual property rights.
We develop a scalable, automated pipeline for collecting high-quality financial news on the Chinese equity market. 
Structured 
parsing and content normalization ensure clean, structured outputs, while resilient error handling with exponential backoff supports stable, large-scale collection. 

Built atop this framework, we construct two multimodal datasets: CSMD 300 and CSMD 50, derived from the CSI 300 and SSE 50 indices, respectively. Covering 300 and 50 representative stocks, these datasets provide comprehensive benchmarks for modeling and analyzing China’s equity market.

The comparison of key attributes among our constructed dataset, CSMD, and previous ones is summarized in Table~\ref{tab:basic information}, where our CSMD exhibits the broadest temporal coverage and recency with data collected from authoritative financial media sources. Then, the next step is to further enhance their quality for financial analysis.

\begin{table}[t]
    \centering
    \caption{Basic information of existing datasets and CSMD}
    \fontsize{8}{15}\selectfont 
    \begin{tabular}{ccccc}
        \hline
        \textbf{Dataset} & \textbf{Country} & \textbf{Stocks} & \textbf{Time Range} & \textbf{Text Sources} \\
        \hline
        ACL18\cite{xu2018stock} & US & 88 & 2014-2015 & Twitter \\
        \hline
        CIKM18\cite{wu2018hybrid}  & US & 38 & 2017-2018 & Twitter \\
        \hline
        BIGDATA22\cite{soun2022accurate}  & US & 50 & 2019-2020 & Twitter \\
        \hline
        CMIN-US\cite{luo2023causality}  & US & 110 & 2018-2021 & Yahoo \\
        \hline
        CMIN-CN\cite{luo2023causality}  & China & 300 & 2018-2021 & Wind \\
        \hline
        \textbf{CSMD 300} & China & 300 & 2021-2024 & STCN \\
        \hline
        \textbf{CSMD 50} & China & 50 & 2021-2024 & STCN \\
        \hline
    \end{tabular}
    \label{tab:basic information}
\end{table}

\begin{figure}[t]
  \includegraphics[width=\linewidth]{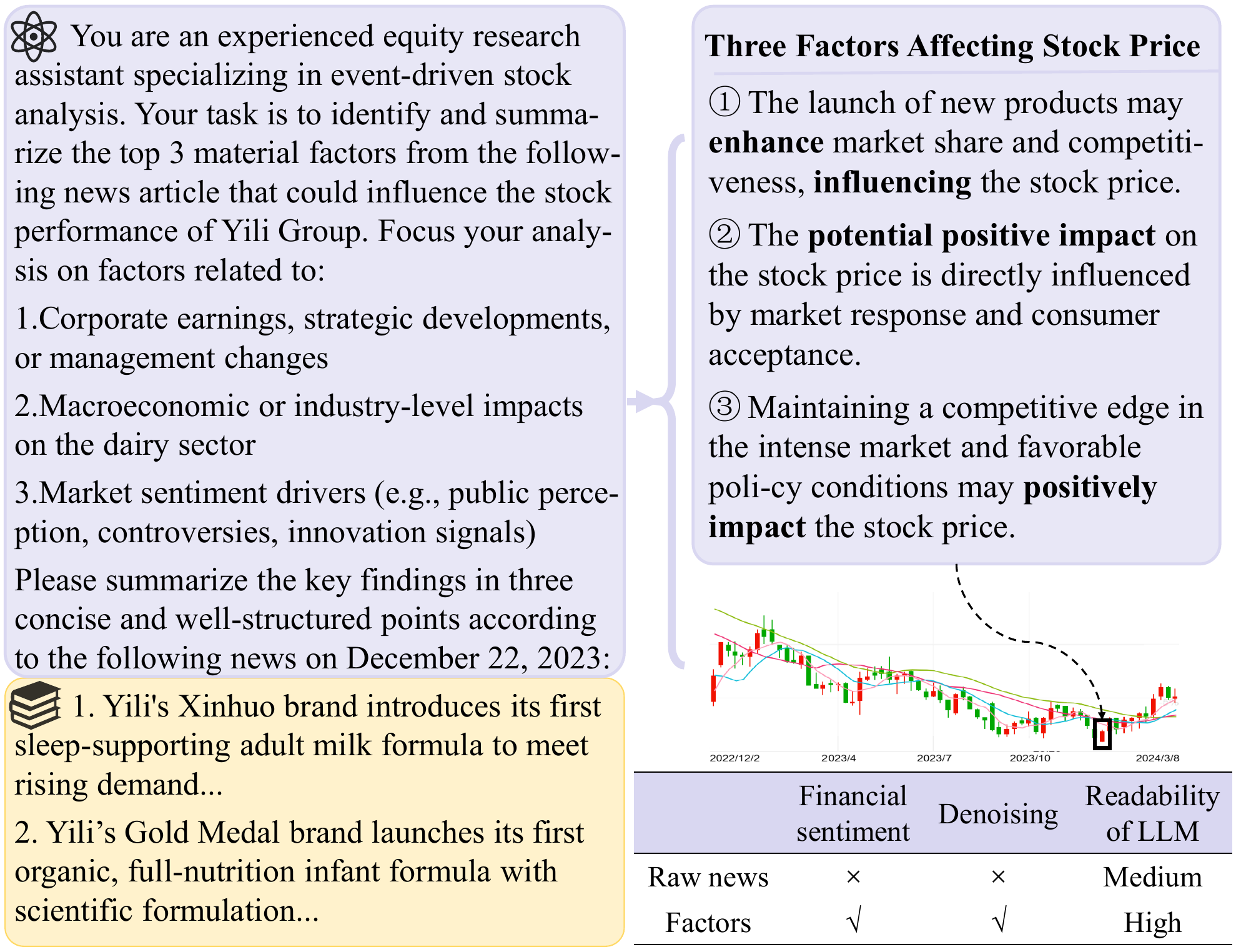}
  \caption{This figure illustrates the application of a LLM to denoise news texts concerning Yili Co., Ltd. from November 22, 2023, yielding factors with high readability and interpretability. The extracted factors are consistent with the stock's actual movement on the subsequent trading day. Note that the original text is in Chinese and has been translated for illustrative purposes.}
  \label{fig:example}
\end{figure}

\subsection{Factor Extraction Enhanced by LLM}

Achieving an accurate forecast of stock movements requires that the extracted stock market impact factors be interpretable, salient, and high-quality.
Thus, after collecting the original financial news data, a second step is to precisely extract the factors from the news texts.
Intuitively, this can be done by integrating domain-specific background knowledge into instructing the LLMs to achieve the automatic selection of the factors~\cite{wang2024llmfactor}.
Specifically, each collected news item is processed through our carefully crafted prompts. 
Then, only the human-readable, explainable, and impactful factors are extracted, which are used to facilitate downstream analysis and improve interpretability with validation and prioritization.
A case is shown in Figure~\ref{fig:example}, illustrating how LLMs can be leveraged to extract impactful news factors from original financial texts. This approach provides enhanced interpretability, readability, and explanatory power compared to raw text, thereby promoting more rigorous and systematic research on stock market analysis.

\begin{table}[t]
    \centering
    \caption{Comparison of data quality among existing datasets.}
    \fontsize{8}{15}\selectfont 
    \begin{tabular}{c@{\hspace{1mm}}c@{\hspace{2mm}}c@{\hspace{1.5mm}}c@{\hspace{1.5mm}}c@{\hspace{1.4mm}}c}
        \hline
        \textbf{Dataset} & \textbf{Denoise} & \textbf{\makecell{Text \\ density}} & \textbf{\makecell{Financial \\ sentiment}} & \textbf{\makecell{Human \\ readability}} & \textbf{\makecell{Readability \\ of LLM}} \\
        \hline
        ACL18 & $\times$ & $\times$ & $\times$ & Low & Low \\
        \hline
        CIKM18 & $\times$ & $\times$ & $\times$ & Low & Low \\
        \hline
        BIGDATA22 & $\times$ & $\times$ & $\times$ & Low & Low \\
        \hline
        CMIN-US & $\checkmark$ & $\checkmark$ & $\times$ & Medium & Medium \\
        \hline
        CMIN-CN & $\checkmark$ & $\checkmark$ & $\times$ & Medium & Medium \\
        \hline
        \textbf{CSMD 300} & $\checkmark$ & $\checkmark$ & $\checkmark$ & High & High\\
        \hline
        \textbf{CSMD 50} & $\checkmark$ & $\checkmark$ & $\checkmark$ & High & High\\
        \hline
    \end{tabular}
    \label{tab:Comparison}
\end{table}

\subsection{Data Quality Validation}
The quality of financial text data used for stock market analysis significantly impacts the downstream model development. 
Thus, the data quality validation is necessary after text data collection and market factor extraction.
To this end, we first categorize the quality validation into five aspects, including denoising, financial sentiment expression, text density, human readability, and LLM readability.
The noisy data would inevitably affect the task performance, e.g., key events and emotions cannot be accurately identified, and the lower readability of humans and LLMs might be a challenge for them in achieving complex analysis~\cite{li2024causalstock}.
Besides, whether the financial sentiment contained in the text data is well extracted is also related to the convenience of downstream tasks.

Then, we conduct manual and automatic evaluations. The manual evaluation is achieved by five human experts from the financial industry to evaluate the readability of the dataset from the perspectives of coherence, relevance, and accuracy.
For the automatic evaluation, we use the MiniLM-L6-v2 model\footnote{https://huggingface.co/cross-encoder/ms-marco-MiniLM-L6-v2} for text sorting and GPT-4 model to score the datasets based on coherence, information content, and topic depth. Comprehensive evaluation details are available in our code repository.
The final comparison results of our CSMD and other existing studies based on the quality of processed financial text are shown in Table~\ref{tab:Comparison}. It can be observed that the CSMD 300 and CSMD 50 datasets exhibit strong performance in denoising, textual richness, financial sentiment alignment, and readability for both human readers and LLMs.

\section{Our LightQuant Framework}
\textbf{LightQuant features.} With the obtained processed datasets, we can conduct various financial downstream tasks on top of available frameworks, e.g., Qlib, an open-source quantitative investment platform that offers end-to-end solutions encompassing data management, feature engineering, model training, evaluation, and backtesting \cite{yang2020qlib}.
However, its complexity and extensive features could be challenging for newcomers to deploy, operate, and debug.
To develop a lightweight and user-friendly framework, we consolidate essential functionalities, including financial data analysis, feature engineering, model training, evaluation, and backtesting by incorporating a plug-and-play modeling module for flexible model integration, and built-in support for fast prototyping and visualization.
The superior performance observed in Section~\ref{experiment} demonstrates the effectiveness of the proposed framework for rapid prototyping and research in financial forecasting.

\begin{figure}[t]
  \centering
  \includegraphics[width=0.95\columnwidth]{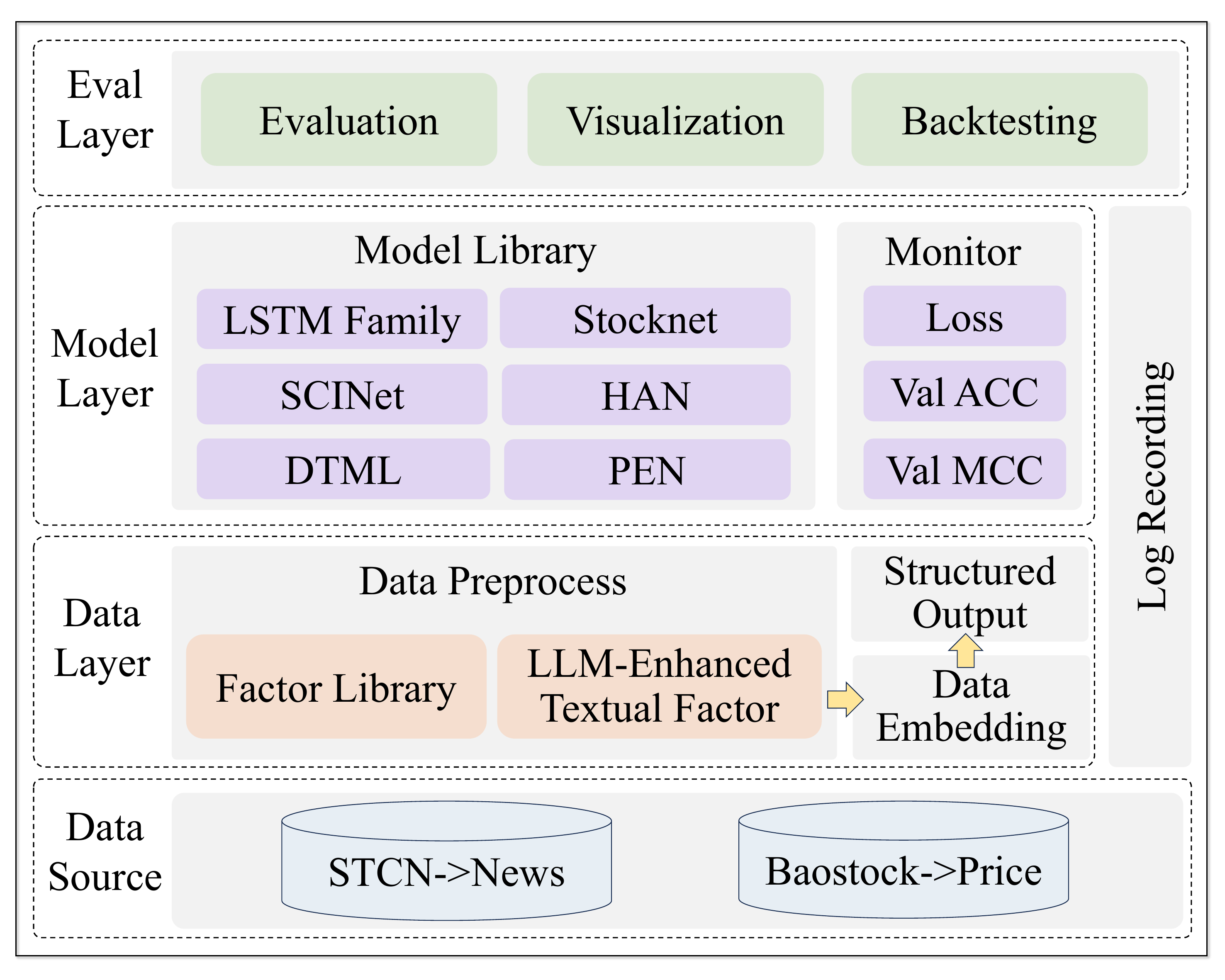}
  \caption{The overall framework of our LightQuant.}
  \label{fig:framework}
\end{figure}

\noindent \textbf{LightQuant Framework.} An overview of the proposed \textit{LightQuant} framework is shown in Figure~\ref{fig:framework}. LightQuant follows a modular architecture consisting of three layers—Data, Model, and Evaluation—designed for flexible adaptation to diverse quantitative research scenarios:

\begin{itemize}
\item \textbf{Data Layer} provides a unified interface for data extraction, processing, and storage across heterogeneous sources. It supports multi-dimensional market data, financial news texts, and integrates feature engineering and factor libraries.
  
\item \textbf{Model Layer} supports the development and integration of various models, including neural architectures for stock prediction. It provides standardized interfaces for model loading, training, and invocation.
  
\item \textbf{Evaluation Layer} delivers evaluation metrics and tools, including back-testing and performance analysis modules, enabling efficient strategy validation and refinement.
\end{itemize}

\section{Experiments} \label{experiment}
\subsection{Experimental Setup}
We conduct downstream task experiments on stock price trend prediction based on our developed LightQuant framework and curated CSMD datasets.
The experimental models are categorized into single-modal and multimodal methods on top of state-of-the-art models.

\noindent \textbf{Single-modal models.} We adopt representative neural models for financial time series forecasting. LSTM \cite{hochreiter1997long} captures sequential dependencies, while BiLSTM \cite{graves2005framewise} enhances contextual learning via bidirectional processing. ALSTM \cite{qin2017dual} introduces attention to focus on critical time steps. Adv-LSTM \cite{miyato2016adversarial} improves robustness through adversarial training. SCINet \cite{liu2022scinet} leverages recursive downsampling and interaction to model complex temporal patterns. DTML \cite{yoo2021accurate}, a transformer-based model, captures multi-level temporal and inter-stock correlations for accurate movement prediction.

\noindent \textbf{Multimodal models.} We further evaluate models that integrate heterogeneous data sources. StockNet \cite{xu2018stock} jointly models textual news and stock prices to exploit cross-modal dependencies. HAN \cite{hu2018listening} applies hierarchical attention to emphasize salient features across modalities. PEN adopts a shared representation framework to efficiently fuse multi-modal inputs, enhancing both prediction accuracy and interpretability. 

\noindent \textbf{Downstream Tasks \& Metrics.} To validate CSMD and LightQuant, we consider two key downstream tasks: stock trend prediction and backtesting. Evaluation employs widely used metrics—accuracy (ACC), Matthews correlation coefficient (MCC), annualized return (ARR), Sharpe ratio (SR), maximum drawdown (MDD), and Calmar ratio (CR)—enabling comprehensive assessment across prediction accuracy, profitability, and risk control.


\begin{table}[t]
    \centering
    \caption{Model comparison on three benchmark datasets. The best results are highlighted in bold, the second best are underlined.}
    \renewcommand{\arraystretch}{1.0} 
    \begin{tabular}{ccccccc}
        \hline
        \textbf{Models} & \multicolumn{2}{c}{\textbf{CSMD 300}} & \multicolumn{2}{c}{\textbf{CSMD 50}} & \multicolumn{2}{c}{\textbf{CMIN-CN}} \\
        \cline{2-3} \cline{4-5} \cline{6-7}
        & \textbf{ACC} &\textbf{MCC} & \textbf{ACC} &\textbf{MCC}  & \textbf{ACC} &\textbf{MCC}  \\
        \hline
        LSTM & 53.37&0.0532&53.61&0.0729&53.28&0.0548 \\
        BiLSTM & 53.70&0.0601&53.98&0.0780&53.46&\underline{0.0571}\\
        ALSTM & 53.62&0.0633&54.21&0.0886&53.35&0.0236 \\
        Adv-LSTM & 54.08&0.0662&54.01&0.0840&53.49&0.0253 \\
        \hline
        SCINet & 53.36&0.0595&53.02&0.0534&53.11&0.0541 \\
        DTML & 54.13&\textbf{0.1478}&54.12&\underline{0.0972}&53.42&0.0334 \\
        StockNet & \textbf{55.47}&0.0916&\textbf{55.11}&0.0746&\textbf{54.53}&0.0450\\
        HAN & \underline{55.00} & \underline{0.0972} & \underline{54.69} & \textbf{0.1002} &53.59&0.0259\\
        PEN &53.27&0.0893 &53.12 & 0.0582 & \underline{53.68}& \textbf{0.0871} \\
        \hline
    \end{tabular}
    \label{tab:performance comparison}
\end{table}

\subsection{Experimental Results}
Table~\ref{tab:performance comparison} reports stock trend prediction results on our CSMD datasets and prior Chinese stock market datasets.
CSMD 300 and CSMD 50 consistently outperform the widely used CMIN-CN across most models, reflecting higher data quality, richer multimodal features, and improved representations from our curation process. These findings highlight the importance of dataset quality in reliable financial modeling and its potential to support robust predictive modeling and advance research in multimodal financial forecasting.


\begin{table}[t]
\caption{The back-testing results on the CSMD 50. Performance here is the average result of 50 stocks. The best results are highlighted in bold, the second best are underlined.}
\centering
\begin{tabular}{l@{\hspace{2.5em}}c@{\hspace{2.5em}}c@{\hspace{2.5em}}c@{\hspace{2.5em}}c}
\toprule
\textbf{Models} & \textbf{ARR} & \textbf{SR} & \textbf{MDD} & \textbf{CR} \\
\midrule
LSTM      & 0.0884 & -0.9203 & 0.0256 & 2.7739 \\
BiLSTM    & 0.0864 & -0.7556 & 0.0250 & 2.8349 \\
ALSTM     & \underline{0.1203} &  \textbf{0.8192} & 0.0308 & 2.9170 \\
Adv-LSTM  & 0.0734 & -1.2998 & 0.0227 & 2.7466 \\
SCINet    & 0.0269  & 0.0627  & \underline{0.0219}& 0.8151 \\
DTML      & 0.0228 & -1.4838  & 0.0597 &0.5469  \\
StockNet  & \textbf{0.1301} & \underline{0.7108}  & 0.0318 & \textbf{3.0182}  \\
HAN       & 0.0764 &  -0.8147 & \textbf{0.0149}  & \underline{2.9763} \\
PEN       & 0.1004  & 0.0271  & 0.0351 & 1.6830 \\
\bottomrule
\end{tabular}
\label{tab:backtest}
\end{table}

Besides, Table~\ref{tab:backtest} summarizes the average performance of baseline models in the backtesting experiment conducted on the CSMD 50 dataset, evaluating their practical trading effectiveness.
StockNet achieves the best overall results, recording the highest ARR and CR. ALSTM obtains the highest Sharpe Ratio, indicating strong risk-adjusted returns. HAN exhibits the lowest MDD. Overall, these results highlight the potential of our dataset to facilitate effective simulation of real-world trading strategies by multimodal models.

\section{Conclusion}
In this paper, we aim to facilitate Chinese stock analysis by curating a multimodal dataset, CSMD, and developing a lightweight, modular framework, LightQuant.
CSMD is constructed through meticulous processing with validated quality compared to existing ones.
Besides, LightQuant framework can support diverse experimental conditions, which provides a user-friendly tool for financial research.
The experimental results demonstrate better efficiency of the LightQuant framework and the effectiveness of using CSMD dataset for developing various models on multimodal financial analytics and stock market forecasting.

\begin{acks}
This work was supported by the Guizhou Provincial Program on Commercialization of Scientific and Technological Achievements (Qiankehezhongyindi [2025] No. 006), the National Natural Science Foundation of China under grant number 62202170, and Seek Data Group, Emoney Inc.
\end{acks}

\section*{GenAI Usage Disclosure}
We utilize GenAI tools such as ChatGPT to optimize and improve spelling and grammar. All core content is independently created by us, and we take full responsibility for the content presented.

\bibliographystyle{ACM-Reference-Format}
\balance
\bibliography{references}


\begin{thebibliography}{24}


\ifx \showCODEN    \undefined \def \showCODEN     #1{\unskip}     \fi
\ifx \showISBNx    \undefined \def \showISBNx     #1{\unskip}     \fi
\ifx \showISBNxiii \undefined \def \showISBNxiii  #1{\unskip}     \fi
\ifx \showISSN     \undefined \def \showISSN      #1{\unskip}     \fi
\ifx \showLCCN     \undefined \def \showLCCN      #1{\unskip}     \fi
\ifx \shownote     \undefined \def \shownote      #1{#1}          \fi
\ifx \showarticletitle \undefined \def \showarticletitle #1{#1}   \fi
\ifx \showURL      \undefined \def \showURL       {\relax}        \fi
\providecommand\bibfield[2]{#2}
\providecommand\bibinfo[2]{#2}
\providecommand\natexlab[1]{#1}
\providecommand\showeprint[2][]{arXiv:#2}

\bibitem[Adam et~al\mbox{.}(2016)]%
        {adam2016stock}
\bibfield{author}{\bibinfo{person}{Klaus Adam}, \bibinfo{person}{Albert Marcet}, {and} \bibinfo{person}{Juan~Pablo Nicolini}.} \bibinfo{year}{2016}\natexlab{}.
\newblock \showarticletitle{Stock market volatility and learning}.
\newblock \bibinfo{journal}{\emph{The Journal of finance}} \bibinfo{volume}{71}, \bibinfo{number}{1} (\bibinfo{year}{2016}), \bibinfo{pages}{33--82}.
\newblock


\bibitem[Antoniou et~al\mbox{.}(1997)]%
        {antoniou1997technical}
\bibfield{author}{\bibinfo{person}{Antonios Antoniou}, \bibinfo{person}{NURAY Ergul}, \bibinfo{person}{PHIL Holmes}, {and} \bibinfo{person}{Richard Priestley}.} \bibinfo{year}{1997}\natexlab{}.
\newblock \showarticletitle{Technical analysis, trading volume and market efficiency: evidence from an emerging market}.
\newblock \bibinfo{journal}{\emph{Applied Financial Economics}} \bibinfo{volume}{7}, \bibinfo{number}{4} (\bibinfo{year}{1997}), \bibinfo{pages}{361--365}.
\newblock


\bibitem[De~Prado(2018)]%
        {de2018advances}
\bibfield{author}{\bibinfo{person}{Marcos~Lopez De~Prado}.} \bibinfo{year}{2018}\natexlab{}.
\newblock \bibinfo{booktitle}{\emph{Advances in financial machine learning}}.
\newblock \bibinfo{publisher}{John Wiley \& Sons}.
\newblock


\bibitem[Graves and Schmidhuber(2005)]%
        {graves2005framewise}
\bibfield{author}{\bibinfo{person}{Alex Graves} {and} \bibinfo{person}{J{\"u}rgen Schmidhuber}.} \bibinfo{year}{2005}\natexlab{}.
\newblock \showarticletitle{Framewise phoneme classification with bidirectional LSTM and other neural network architectures}.
\newblock \bibinfo{journal}{\emph{Neural networks}} \bibinfo{volume}{18}, \bibinfo{number}{5-6} (\bibinfo{year}{2005}), \bibinfo{pages}{602--610}.
\newblock


\bibitem[Hochreiter and Schmidhuber(1997)]%
        {hochreiter1997long}
\bibfield{author}{\bibinfo{person}{Sepp Hochreiter} {and} \bibinfo{person}{J{\"u}rgen Schmidhuber}.} \bibinfo{year}{1997}\natexlab{}.
\newblock \showarticletitle{Long short-term memory}.
\newblock \bibinfo{journal}{\emph{Neural computation}} \bibinfo{volume}{9}, \bibinfo{number}{8} (\bibinfo{year}{1997}), \bibinfo{pages}{1735--1780}.
\newblock


\bibitem[Hu et~al\mbox{.}(2018)]%
        {hu2018listening}
\bibfield{author}{\bibinfo{person}{Ziniu Hu}, \bibinfo{person}{Weiqing Liu}, \bibinfo{person}{Jiang Bian}, \bibinfo{person}{Xuanzhe Liu}, {and} \bibinfo{person}{Tie-Yan Liu}.} \bibinfo{year}{2018}\natexlab{}.
\newblock \showarticletitle{Listening to chaotic whispers: A deep learning framework for news-oriented stock trend prediction}. In \bibinfo{booktitle}{\emph{Proceedings of the eleventh ACM international conference on web search and data mining}}. \bibinfo{pages}{261--269}.
\newblock


\bibitem[Huang et~al\mbox{.}(2024)]%
        {huang2024survey}
\bibfield{author}{\bibinfo{person}{Kaiyu Huang}, \bibinfo{person}{Fengran Mo}, \bibinfo{person}{Xinyu Zhang}, \bibinfo{person}{Hongliang Li}, \bibinfo{person}{You Li}, \bibinfo{person}{Yuanchi Zhang}, \bibinfo{person}{Weijian Yi}, \bibinfo{person}{Yulong Mao}, \bibinfo{person}{Jinchen Liu}, \bibinfo{person}{Yuzhuang Xu}, {et~al\mbox{.}}} \bibinfo{year}{2024}\natexlab{}.
\newblock \showarticletitle{A survey on large language models with multilingualism: Recent advances and new frontiers}.
\newblock \bibinfo{journal}{\emph{arXiv preprint arXiv:2405.10936}} (\bibinfo{year}{2024}).
\newblock


\bibitem[Li et~al\mbox{.}(2024)]%
        {li2024causalstock}
\bibfield{author}{\bibinfo{person}{Shuqi Li}, \bibinfo{person}{Yuebo Sun}, \bibinfo{person}{Yuxin Lin}, \bibinfo{person}{Xin Gao}, \bibinfo{person}{Shuo Shang}, {and} \bibinfo{person}{Rui Yan}.} \bibinfo{year}{2024}\natexlab{}.
\newblock \showarticletitle{CausalStock: Deep end-to-end causal discovery for news-driven multi-stock movement prediction}.
\newblock \bibinfo{journal}{\emph{Advances in Neural Information Processing Systems}}  \bibinfo{volume}{37} (\bibinfo{year}{2024}), \bibinfo{pages}{47432--47454}.
\newblock


\bibitem[Liu et~al\mbox{.}(2022)]%
        {liu2022scinet}
\bibfield{author}{\bibinfo{person}{Minhao Liu}, \bibinfo{person}{Ailing Zeng}, \bibinfo{person}{Muxi Chen}, \bibinfo{person}{Zhijian Xu}, \bibinfo{person}{Qiuxia Lai}, \bibinfo{person}{Lingna Ma}, {and} \bibinfo{person}{Qiang Xu}.} \bibinfo{year}{2022}\natexlab{}.
\newblock \showarticletitle{Scinet: Time series modeling and forecasting with sample convolution and interaction}.
\newblock \bibinfo{journal}{\emph{Advances in Neural Information Processing Systems}}  \bibinfo{volume}{35} (\bibinfo{year}{2022}), \bibinfo{pages}{5816--5828}.
\newblock


\bibitem[Luo et~al\mbox{.}(2023)]%
        {luo2023causality}
\bibfield{author}{\bibinfo{person}{Di Luo}, \bibinfo{person}{Weiheng Liao}, \bibinfo{person}{Shuqi Li}, \bibinfo{person}{Xin Cheng}, {and} \bibinfo{person}{Rui Yan}.} \bibinfo{year}{2023}\natexlab{}.
\newblock \showarticletitle{Causality-guided multi-memory interaction network for multivariate stock price movement prediction}. In \bibinfo{booktitle}{\emph{Proceedings of the 61st Annual Meeting of the Association for Computational Linguistics (Volume 1: Long Papers)}}. \bibinfo{pages}{12164--12176}.
\newblock


\bibitem[Mao et~al\mbox{.}(2023)]%
        {mao2023discovering}
\bibfield{author}{\bibinfo{person}{Rui Mao}, \bibinfo{person}{Kelvin Du}, \bibinfo{person}{Yu Ma}, \bibinfo{person}{Luyao Zhu}, {and} \bibinfo{person}{Erik Cambria}.} \bibinfo{year}{2023}\natexlab{}.
\newblock \showarticletitle{Discovering the cognition behind language: Financial metaphor analysis with MetaPro}. In \bibinfo{booktitle}{\emph{2023 IEEE International Conference on Data Mining (ICDM)}}. IEEE, \bibinfo{pages}{1211--1216}.
\newblock


\bibitem[Miyato et~al\mbox{.}(2016)]%
        {miyato2016adversarial}
\bibfield{author}{\bibinfo{person}{Takeru Miyato}, \bibinfo{person}{Andrew~M Dai}, {and} \bibinfo{person}{Ian Goodfellow}.} \bibinfo{year}{2016}\natexlab{}.
\newblock \showarticletitle{Adversarial training methods for semi-supervised text classification}.
\newblock \bibinfo{journal}{\emph{arXiv preprint arXiv:1605.07725}} (\bibinfo{year}{2016}).
\newblock


\bibitem[Nassirtoussi(2015)]%
        {Nassirtoussi2015AMD}
\bibfield{author}{\bibinfo{person}{Arman~Khadjeh Nassirtoussi}.} \bibinfo{year}{2015}\natexlab{}.
\newblock \showarticletitle{A multi-layer dimension reduction algorithm for text mining of news in forex / Arman Khadjeh Nassirtoussi}.
\newblock
\urldef\tempurl%
\url{https://api.semanticscholar.org/CorpusID:61883046}
\showURL{%
\tempurl}


\bibitem[Nassirtoussi et~al\mbox{.}(2015)]%
        {nassirtoussi2015text}
\bibfield{author}{\bibinfo{person}{Arman~Khadjeh Nassirtoussi}, \bibinfo{person}{Saeed Aghabozorgi}, \bibinfo{person}{Teh~Ying Wah}, {and} \bibinfo{person}{David Chek~Ling Ngo}.} \bibinfo{year}{2015}\natexlab{}.
\newblock \showarticletitle{Text mining of news-headlines for FOREX market prediction: A Multi-layer Dimension Reduction Algorithm with semantics and sentiment}.
\newblock \bibinfo{journal}{\emph{Expert Systems with Applications}} \bibinfo{volume}{42}, \bibinfo{number}{1} (\bibinfo{year}{2015}), \bibinfo{pages}{306--324}.
\newblock


\bibitem[Naveed et~al\mbox{.}(2023)]%
        {naveed2023comprehensive}
\bibfield{author}{\bibinfo{person}{Humza Naveed}, \bibinfo{person}{Asad~Ullah Khan}, \bibinfo{person}{Shi Qiu}, \bibinfo{person}{Muhammad Saqib}, \bibinfo{person}{Saeed Anwar}, \bibinfo{person}{Muhammad Usman}, \bibinfo{person}{Naveed Akhtar}, \bibinfo{person}{Nick Barnes}, {and} \bibinfo{person}{Ajmal Mian}.} \bibinfo{year}{2023}\natexlab{}.
\newblock \showarticletitle{A comprehensive overview of large language models}.
\newblock \bibinfo{journal}{\emph{arXiv preprint arXiv:2307.06435}} (\bibinfo{year}{2023}).
\newblock


\bibitem[Qin et~al\mbox{.}(2017)]%
        {qin2017dual}
\bibfield{author}{\bibinfo{person}{Yao Qin}, \bibinfo{person}{Dongjin Song}, \bibinfo{person}{Haifeng Chen}, \bibinfo{person}{Wei Cheng}, \bibinfo{person}{Guofei Jiang}, {and} \bibinfo{person}{Garrison Cottrell}.} \bibinfo{year}{2017}\natexlab{}.
\newblock \showarticletitle{A dual-stage attention-based recurrent neural network for time series prediction}.
\newblock \bibinfo{journal}{\emph{arXiv preprint arXiv:1704.02971}} (\bibinfo{year}{2017}).
\newblock


\bibitem[Soun et~al\mbox{.}(2022)]%
        {soun2022accurate}
\bibfield{author}{\bibinfo{person}{Yejun Soun}, \bibinfo{person}{Jaemin Yoo}, \bibinfo{person}{Minyong Cho}, \bibinfo{person}{Jihyeong Jeon}, {and} \bibinfo{person}{U Kang}.} \bibinfo{year}{2022}\natexlab{}.
\newblock \showarticletitle{Accurate stock movement prediction with self-supervised learning from sparse noisy tweets}. In \bibinfo{booktitle}{\emph{2022 IEEE International Conference on Big Data (Big Data)}}. IEEE, \bibinfo{pages}{1691--1700}.
\newblock


\bibitem[Tang et~al\mbox{.}(2022)]%
        {tang2022survey}
\bibfield{author}{\bibinfo{person}{Yajiao Tang}, \bibinfo{person}{Zhenyu Song}, \bibinfo{person}{Yulin Zhu}, \bibinfo{person}{Huaiyu Yuan}, \bibinfo{person}{Maozhang Hou}, \bibinfo{person}{Junkai Ji}, \bibinfo{person}{Cheng Tang}, {and} \bibinfo{person}{Jianqiang Li}.} \bibinfo{year}{2022}\natexlab{}.
\newblock \showarticletitle{A survey on machine learning models for financial time series forecasting}.
\newblock \bibinfo{journal}{\emph{Neurocomputing}}  \bibinfo{volume}{512} (\bibinfo{year}{2022}), \bibinfo{pages}{363--380}.
\newblock


\bibitem[Wang et~al\mbox{.}(2024)]%
        {wang2024llmfactor}
\bibfield{author}{\bibinfo{person}{Meiyun Wang}, \bibinfo{person}{Kiyoshi Izumi}, {and} \bibinfo{person}{Hiroki Sakaji}.} \bibinfo{year}{2024}\natexlab{}.
\newblock \showarticletitle{LLMFactor: Extracting profitable factors through prompts for explainable stock movement prediction}.
\newblock \bibinfo{journal}{\emph{arXiv preprint arXiv:2406.10811}} (\bibinfo{year}{2024}).
\newblock


\bibitem[Wang et~al\mbox{.}(2025)]%
        {wang2025finsage}
\bibfield{author}{\bibinfo{person}{Xinyu Wang}, \bibinfo{person}{Jijun Chi}, \bibinfo{person}{Zhenghan Tai}, \bibinfo{person}{Tung Sum~Thomas Kwok}, \bibinfo{person}{Muzhi Li}, \bibinfo{person}{Zhuhong Li}, \bibinfo{person}{Hailin He}, \bibinfo{person}{Yuchen Hua}, \bibinfo{person}{Peng Lu}, \bibinfo{person}{Suyuchen Wang}, {et~al\mbox{.}}} \bibinfo{year}{2025}\natexlab{}.
\newblock \showarticletitle{Finsage: A multi-aspect rag system for financial filings question answering}.
\newblock \bibinfo{journal}{\emph{arXiv preprint arXiv:2504.14493}} (\bibinfo{year}{2025}).
\newblock


\bibitem[Wu et~al\mbox{.}(2018)]%
        {wu2018hybrid}
\bibfield{author}{\bibinfo{person}{Huizhe Wu}, \bibinfo{person}{Wei Zhang}, \bibinfo{person}{Weiwei Shen}, {and} \bibinfo{person}{Jun Wang}.} \bibinfo{year}{2018}\natexlab{}.
\newblock \showarticletitle{Hybrid deep sequential modeling for social text-driven stock prediction}. In \bibinfo{booktitle}{\emph{Proceedings of the 27th ACM international conference on information and knowledge management}}. \bibinfo{pages}{1627--1630}.
\newblock


\bibitem[Xu and Cohen(2018)]%
        {xu2018stock}
\bibfield{author}{\bibinfo{person}{Yumo Xu} {and} \bibinfo{person}{Shay~B Cohen}.} \bibinfo{year}{2018}\natexlab{}.
\newblock \showarticletitle{Stock movement prediction from tweets and historical prices}. In \bibinfo{booktitle}{\emph{Proceedings of the 56th Annual Meeting of the Association for Computational Linguistics (Volume 1: Long Papers)}}. \bibinfo{pages}{1970--1979}.
\newblock


\bibitem[Yang et~al\mbox{.}(2020)]%
        {yang2020qlib}
\bibfield{author}{\bibinfo{person}{Xiao Yang}, \bibinfo{person}{Weiqing Liu}, \bibinfo{person}{Dong Zhou}, \bibinfo{person}{Jiang Bian}, {and} \bibinfo{person}{Tie-Yan Liu}.} \bibinfo{year}{2020}\natexlab{}.
\newblock \showarticletitle{Qlib: An ai-oriented quantitative investment platform}.
\newblock \bibinfo{journal}{\emph{arXiv preprint arXiv:2009.11189}} (\bibinfo{year}{2020}).
\newblock


\bibitem[Yoo et~al\mbox{.}(2021)]%
        {yoo2021accurate}
\bibfield{author}{\bibinfo{person}{Jaemin Yoo}, \bibinfo{person}{Yejun Soun}, \bibinfo{person}{Yong-chan Park}, {and} \bibinfo{person}{U Kang}.} \bibinfo{year}{2021}\natexlab{}.
\newblock \showarticletitle{Accurate multivariate stock movement prediction via data-axis transformer with multi-level contexts}. In \bibinfo{booktitle}{\emph{Proceedings of the 27th ACM SIGKDD Conference on Knowledge Discovery \& Data Mining}}. \bibinfo{pages}{2037--2045}.
\newblock


\end{thebibliography}


\end{document}